\documentclass[12pt]{article}

\usepackage{amsmath}
\usepackage{esint}                   

\usepackage[latin1]{inputenc}
\usepackage{pstricks}
\usepackage{cancel}
\usepackage[textwidth= 17.5cm, height= 24cm]{geometry}
\usepackage{mathrsfs}
\usepackage[T1]{fontenc}
\usepackage{mathpazo}
\usepackage{setspace}
\usepackage{amsfonts}
\usepackage{amssymb}
\usepackage{amsmath}
\usepackage{epsfig}
\usepackage{latexsym}
\usepackage{color}
\usepackage{graphicx}
\usepackage{nicefrac}
\usepackage[latin1]{inputenc}
\usepackage{slashed}
\usepackage{multirow}
\usepackage{comment}
\usepackage{soul}
\usepackage{hyperref}
\usepackage[nosort]{cite}
\usepackage{datetime}
\usepackage{braket}

\usepackage[titletoc]{appendix}

\def\hybrid{\topmargin -30pt    \oddsidemargin 0pt 
        \headheight 0pt \headsep 0pt
        \textwidth 6.25in       
        \textheight 9.5in       
        \marginparwidth .875in
        \parskip 5pt plus 1pt   \jot = 1.5ex}

\hybrid   

\def\baselinestretch{1.2}

\catcode`\@=11

\def\marginnote#1{}
\newcount\hour
\newcount\minute
\newtoks\amorpm
\hour=\time\divide\hour by60
\minute=\time{\multiply\hour by60 \global\advance\minute by-\hour}
\edef\standardtime{{\ifnum\hour<12 \global\amorpm={am}%
        \else\global\amorpm={pm}\advance\hour by-12 \fi
        \ifnum\hour=0 \hour=12 \fi
        \number\hour:\ifnum\minute<10 0\fi\number\minute\the\amorpm}}
\edef\militarytime{\number\hour:\ifnum\minute<10 0\fi\number\minute}

\def\draftlabel#1{{\@bsphack\if@filesw {\let\thepage\relax
   \xdef\@gtempa{\write\@auxout{\string
      \newlabel{#1}{{\@currentlabel}{\thepage}}}}}\@gtempa
   \if@nobreak \ifvmode\nobreak\fi\fi\fi\@esphack}
        \gdef\@eqnlabel{#1}}
\def\@eqnlabel{}
\def\@vacuum{}
\def\draftmarginnote#1{\marginpar{\raggedright\scriptsize\tt#1}}

\def\draft{\oddsidemargin -.5truein
        \def\@oddfoot{\sl preliminary draft \hfil
        \rm\thepage\hfil\sl\today\quad\militarytime}
        \let\@evenfoot\@oddfoot \overfullrule 3pt
        \let\label=\draftlabel
        \let\marginnote=\draftmarginnote
   \def\@eqnnum{(\theequation)\rlap{\kern\marginparsep\tt\@eqnlabel}%
\global\let\@eqnlabel\@vacuum}  }

\def\draft2{
        \def\@oddfoot{\sl preliminary draft \hfil
        \rm\thepage\hfil\sl\today\quad\militarytime}
        \let\@evenfoot\@oddfoot \overfullrule 3pt
        \let\label=\draftlabel
        \let\marginnote=\draftmarginnote
   \def\@eqnnum{(\theequation)\rlap{\kern\marginparsep\tt\@eqnlabel}%
\global\let\@eqnlabel\@vacuum}  }

\def\preprint{\twocolumn\sloppy\flushbottom\parindent 2em
        \leftmargini 2em\leftmarginv .5em\leftmarginvi .5em
        \oddsidemargin -.5in    \evensidemargin -.5in
        \columnsep .4in \footheight 0pt
        \textwidth 10.in        \topmargin  -.4in
        \headheight 12pt \topskip .4in
        \textheight 6.9in \footskip 0pt
        \def\@oddhead{\thepage\hfil\addtocounter{page}{1}\thepage}
        \let\@evenhead\@oddhead \def\@oddfoot{} \def\@evenfoot{} }

\def\numberbysection{\@addtoreset{equation}{section}
        \def\theequation{\thesection.\arabic{equation}}}

\def\underline#1{\relax\ifmmode\@@underline#1\else
        $\@@underline{\hbox{#1}}$\relax\fi}

\def\titlepage{\@restonecolfalse\if@twocolumn\@restonecoltrue\onecolumn
     \else \newpage \fi \thispagestyle{empty}\c@page\z@
        \def\thefootnote{\fnsymbol{footnote}} }

\def\endtitlepage{\if@restonecol\twocolumn \else \newpage \fi
        \def\thefootnote{\arabic{footnote}}
        \setcounter{footnote}{0}}  

\catcode`@=12
\relax

\def\figcap{\section*{Figure Captions\markboth
        {FIGURECAPTIONS}{FIGURECAPTIONS}}\list
        {Figure \arabic{enumi}:\hfill}{\settowidth\labelwidth{Figure
999:}
        \leftmargin\labelwidth
        \advance\leftmargin\labelsep\usecounter{enumi}}}
 \relax
\def\tablecap{\section*{Table Captions\markboth
        {TABLECAPTIONS}{TABLECAPTIONS}}\list
        {Table \arabic{enumi}:\hfill}{\settowidth\labelwidth{Table
999:}
        \leftmargin\labelwidth
        \advance\leftmargin\labelsep\usecounter{enumi}}}
 \relax
\def\reflist{\section*{References\markboth
        {REFLIST}{REFLIST}}\list
        {[\arabic{enumi}]\hfill}{\settowidth\labelwidth{[999]}
        \leftmargin\labelwidth
        \advance\leftmargin\labelsep\usecounter{enumi}}}
 \relax
\makeatletter
\newcounter{pubctr}
\def\publist{\@ifnextchar[{\@publist}{\@@publist}}
\def\@publist[#1]{\list
        {[\arabic{pubctr}]\hfill}{\settowidth\labelwidth{[999]}
        \leftmargin\labelwidth
        \advance\leftmargin\labelsep
        \@nmbrlisttrue\def\@listctr{pubctr}
        \setcounter{pubctr}{#1}\addtocounter{pubctr}{-1}}}
\def\@@publist{\list
        {[\arabic{pubctr}]\hfill}{\settowidth\labelwidth{[999]}
        \leftmargin\labelwidth
        \advance\leftmargin\labelsep
        \@nmbrlisttrue\def\@listctr{pubctr}}}
 \relax
\makeatother

\def\be{\begin{equation}}
\def\ee{\end{equation}}
\def\ba{\begin{eqnarray}}
\def\ea{\end{eqnarray}}

\def\r{\rho}
\def\a{\alpha}

\def\b{\beta}

\def\d{\delta}

\def\m{\mu}
\def\n{\nu}
\def\om{\omega}

\def\l{\lambda}

\def\s{\sigma}

\def\cA{{\cal A}}

\def\cD{{\cal D}}
\def\cE{{\cal E}}

\def\cT{{\cal T}}

\def\cM{{\cal M}}

\def\no{\noindent}

\def\qq{\qquad}

\def\IR{\relax{\rm I\kern-.18em R}}

\def\inv{^{\raise.0ex\hbox{${\scriptscriptstyle -}$}\kern-.05em 1}}

\def \ov {\over}

\begin{document}


\renewcommand{\theequation}{\thesection.\arabic{equation}}
\csname @addtoreset\endcsname{equation}{section}

\begin{titlepage}
\begin{center}

\phantom{xx}
\vskip -0.3in
 
 Nucl.Phys. {\bf B953} (2020) 114960  \hfill\break

\vskip 0.3in

{\large \bf $AdS$ solutions and $\l$-deformations
}

\vskip 0.5in

{\bf Georgios Itsios}${}^{1a}$\phantom{x} and \phantom{x} {\bf Konstantinos Sfetsos}${}^{2b}$ 

\vskip 0.1in

${}^1$ Instituto de F\'isica Te\'orica, UNESP-Universidade Estadual Paulista, \\
R. Dr. Bento T. Ferraz 271, Bl. II, Sao Paulo 01140-070, SP, Brazil.\\[1cm]

\vskip -0.15 in

${}^2$ Department of Nuclear and Particle Physics,\\
Faculty of Physics, National and Kapodistrian University of Athens,\\
Athens 15784, Greece

\vskip .2in


\end{center}

\vskip .4in

\centerline{\bf Abstract}

\vskip .08in

\no
We elevate $\l$-deformed $\s$-models into full type-II supergravity backgrounds. We construct
several solutions which contain undeformed $AdS_n$ spaces, with $n=2,3,4$ and $6$,
as an integrable part. In that respect, our examples are the first in the literature in this context and bring $\l$-deformations in contact with the AdS/CFT correspondence.
The geometries are supported by appropriate dilaton and RR-fields. 
Most of the solutions admit non-Abelian T-dual limits.

\vfill
\no
 {\footnotesize
$^a$gitsios@gmail.com,
$^b$ksfetsos@phys.uoa.gr
}

\end{titlepage}
\vfill
\eject



\def\baselinestretch{1.2}
\baselineskip 20 pt

\newcommand{\eqn}[1]{(\ref{#1})}

\tableofcontents


\section{Introduction}

In recent years there has been a lot of activity in constructing classes of two-dimensional quantum field theories having a
group theoretical basis, remarkable properties and controllable behavior. One such class comprises theories that generically come under the name of $\l$-deformations. In the prototype example \cite{Sfetsos:2013wia} the model is characterized by
a conformal field theory (CFT) having a current algebra symmetry at level $k$ and a Lagrangian realization in terms
of a Wess-Zumino-Witten (WZW) model \cite{Witten:1983ar}, as well as by a matrix $\l$. For small entries of this matrix it is just the action of the WZW theory perturbed by current bilinears $\l_{ab} J^a_+J_-^b$. The full action is non-linear in $\l$, effectively takes into account all loop quantum effects arising  from the self-interaction of the chiral and anti-chiral currents of the original
unperturbed WZW theory and is integrable for $\l$ proportional to the identity  \cite{Sfetsos:2013wia}.
Since then, this construction has been extended to cover cases with more than one
current algebra theories, mutually and/or  self-interacting \cite{Georgiou:2016zyo,Georgiou:2017jfi,Georgiou:2018hpd,Georgiou:2018gpe}.
These models have a very rich structure and their RG flow possesses several fixed points.
The use of non-trivial outer automorphisms in this context was put forward  for the case of a single group in \cite{Driezen:2019ykp}.
A major reason that the $\l$-deformed models have attracted attention is due to the fact that they are integrable for specific choices of the deformation matrix $\l$. Such cases exist first for isotropic deformation matrices
\cite{Sfetsos:2013wia,Georgiou:2017jfi,Georgiou:2018hpd,Georgiou:2018gpe}
(for the $SU(2)$ group case, integrability was shown in \cite{Balog:1993es})
as well as for some anisotropic model cases \cite{Sfetsos:2014lla,Sfetsos:2015nya} and for subclasses of the models
 of \cite{Georgiou:2018hpd,Georgiou:2018gpe}. Integrable deformations based on cosets having a Lagrangian
description in terms of gauged WZW models, symmetric and semi-symmetric spaces have also been constructed in \cite{Sfetsos:2013wia}, \cite{Hollowood:2014rla,Hollowood:2014qma} and \cite{Sfetsos:2017sep}.

\no
In parallel to the above, a class of integrable $\s$-models was introduced in \cite{Klimcik:2002zj,Klimcik:2008eq,Klimcik:2014}
and \cite{Delduc:2013fga,Delduc:2013qra,Arutyunov:2013ega} for group and coset spaces, respectively. These are ultimately
connected to the Principal Chiral Model (PCM) for groups and cosets which are integrable.
There is a relation between $\l$-deformations and $\eta$-deformations for group and coset spaces as was discussed in  \cite{Vicedo:2015pna,Hoare:2015gda}, \cite{Sfetsos:2015nya,Klimcik:2015gba,Klimcik:2016rov,Hoare:2018ebg}. In particular,
in these works it was shown that for the isotropic case the  $\l$-deformed model is related via Poisson-Lie T-duality \cite{KS95a} and appropriate analytic continuations to the $\eta$-deformed model. For a comprehensible review concerning certain aspects of $\l$- and $\eta$-deformations see 
\cite{Thompson:2019ipl}.

Having a $\s$-model, let alone an integrable one, it is very important to elevate it into
a full supergravity solution.
That is, support the metric and the antisymmetric tensor fields of the two-dimensional $\s$-model with the appropriate dilaton and Ramond-Ramond (RR) fields so that the equations of motion are satisfied.
Nevertheless, this is in general a challenging task to perform.
Several examples in this direction have been already presented in the literature \cite{Sfetsos:2014cea,Demulder:2015lva,Hoare:2015gda,Borsato:2016zcf,Chervonyi:2016ajp,
Borsato:2016ose,Hoare:2015wia,Lunin:2014tsa,Hoare:2018ngg,Seibold:2019dvf}. A common, characteristic of these works, no matter whether
it concerned $\l$- or $\eta$-deformations, is that the entire non-trivial part of the space-time is
deformed. Hence, undeformed spaces like $AdS$ were not found so far as part of full
supergravity solutions.
This is the task we  successfully  overtake in the present paper by constructing type-II
supergravity solutions containing both $\l$-deformed and $AdS$  spaces. Therefore, they 
can be used in the context of the AdS/CFT correspondence.

The outline of this paper is as follows: In Sec. \ref{Sec2} we review the $\l$-deformed models based on the coset CFTs $SU(2)/U(1)$, $SL(2,\mathbb{R})/U(1)$ and $SL(2,\mathbb{R})/SO(1,1)$. In Sec. \ref{Sec3} we continue by
 constructing solutions of the type-IIA and type-IIB supergravity theories with geometry $\cM_6 \times CS^2_\l \times CH_{2,\l}$.\footnote{The presence of the symbol $\l$ in the subscript indicates that this is a deformed space. The precise definitions
 will be given later in the text.} We found that the corresponding geometries admit $AdS_2$ and $AdS_3$ factors, respectively. In the same spirit, in Sec. \ref{Sec4} we present various type-IIB backgrounds with geometry $\cM_8 \times CS^2_\l$ which include $AdS_2$, $AdS_4$ and $AdS_6$ factors. In Sec. \ref{Sec5} we find solutions of the type-IIB supergravity, whose metric is of the form $\cM_8 \times CH_{2,\l}$ and includes $AdS_2$ factors. Next, in Sec. \ref{Sec6}, after reviewing the deformed CFTs on $SO(4)/SO(3)$ and $SO(3,1)/SO(2,1)$ we construct a type-IIB background
 with geometry $\cM_4 \times CS^3_\l \times CH_{3,\l}$. This solution allows for an $AdS_2$ factor.
 Finally, Sec. \ref{Sec7} contains our concluding remarks. We complete the presentation with two appendices. In Appendix \ref{AppA} we give an account of the equations of motion of the type-IIA and type-IIB supergravities, while in Appendix \ref{AppB} we discuss the non-Abelian T-duality limit of the deformed models and explicitly perform it to a solution of Sec. \ref{Sec3}.

\section{Two-dimensional $\l$-deformed models}
\label{Sec2}

Here we review the $\l$-deformed models based on the coset $SU(2)/U(1)$ CFT
constructed in \cite{Sfetsos:2013wia}. Then by analytic continuation we obtain the corresponding
$\l$-deformed models based on the $SL(2,\mathbb{R})/U(1)$ and $SL(2,\mathbb{R})/SO(1,1)$ exact coset CFTs.

\subsection{$\l$-deformed model based on the coset $SU(2)/U(1)$}

The geometry of the $\l$-deformed $SU(2)/U(1)$ is given by  \cite{Sfetsos:2013wia}
\begin{equation}
\label{metric1}
 ds^2 = k \frac{1 - \l}{ 1 + \l} \big(   d\b^2 + \cot^2 \b \, d\a^2 \big) + \frac{4 k \l}{1 - \l^2} \big(  \sin\a \, \cot\b \, d\a + \cos\a \, d\b  \big)^2 \, ,
\end{equation}
When $\l = 0$ one recovers the undeformed geometry corresponding to the $SU(2)/U(1)$ exact CFT
\cite{Bardacki:1990wj}
\begin{equation}
 ds^2 = k \big(   d\b^2 + \cot^2 \b \, d\a^2 \big) \ .
\end{equation}
The geometry \eqn{metric1} is supported by a non-trivial dilaton arising from integrating out gauge fields in the construction \cite{Sfetsos:2013wia}
\begin{equation}
\label{DilatonS2La}
 \Phi = - \ln \sin\b \, ,
\end{equation}
which we note is $\l$-independent.
We will denote the background corresponding to \eqn{metric1} and \eqn{DilatonS2La} as $CS^2_\l$.

\no
It is convenient to define the following combinations of constants
\begin{equation}
 \l_\pm := \sqrt{k \frac{1 \pm \l}{1 \mp \l}} \ ,\qq  \m := \frac{4 \l}{k (1 - \l^2)} \ , \qquad \n := \frac{4}{k} \frac{1 + \l^2}{1 - \l^2} \  .
\label{lpm}
\end{equation}
It is also useful to express the deformed geometry using the following frame
\footnote{
\label{FrameSimp}
Notice that the frame can be written as $\mathfrak{e}^1 = e^{\Phi} dx_1$ and $\mathfrak{e}^2 = e^{\Phi} dx_2$ where $x_1 = \l_- \sin\a \cos\b$ and $x_2 = - \l_+ \cos\a \cos\b$. As a result the metric is conformally flat and the dilaton plays the r\^ole of the conformal factor, i.e. $ds^2 = e^{2 \Phi} \big(  dx_1^2 + dx_2^2 \big)$.
In these coordinates the dilaton becomes  $e^{-2\Phi}=1-x_1^2/\l_-^2 - x_2^2/\l_+^2$.
}
\begin{equation}
\label{FrameS2La}
 \mathfrak{e}^1 = \l_- \big(  \cos\a \, \cot\b \, d\a - \sin\a \, d\b  \big) \ , \qquad \mathfrak{e}^2 = \l_+ \big(  \sin\a \, \cot\b \, d\a + \cos\a \, d\b  \big) \, .
\end{equation}

\no
The background  $CS^2_\l$ is invariant under  the following symmetries \cite{Itsios:2014lca}
\be
\label{sym1}
\l\to \l^{-1}\ ,\qq k\to -k\  ,
\ee
which leaves the above frames invariant. There is an additional symmetry
that has been unnoticed so far in the literature, acting as
\be
\label{sym2}
\l\to -\l \ ,\qq \a\to \a+{\pi \ov 2}\ ,\qq \b\to -\b\ .
\ee
On the frame it acts as $ \mathfrak{e}^1\to  \mathfrak{e}^2$ and $ \mathfrak{e}^2\to - \mathfrak{e}^1$, leaving
of course the metric invariant.

\no
The dilaton beta function is
\begin{equation}
\label{dilsul}
 \b^{\Phi}_{CS^2_\l} = R + 4 \nabla^2 \Phi - 4 \big(  \partial \Phi \big)^2 = \n \, .
\end{equation}
Another useful combination that plays significant r\^ole in the following sections is
\begin{equation}
\label{HDJ1}
 R_{ab} + 2 \nabla_a \nabla_b \Phi = - \m \, \eta_{ab} \, ,
\end{equation}
where $\eta_{ab}$ is the Minkowski metric with signature $(-,+)$.
This will be instrumental in constructing type-II supergravity
solutions in which the background  $CS^2_\l$ will be an integrable part.
Note also that \eqn{HDJ1} is frame-dependent since the
right hand side is proportional to $\eta_{ab}$ even though the metric is of Euclidean signature. That implies that
an $SO(2)$ rotation of the frames \eqn{FrameS2La} will not leave \eqn{HDJ1} invariant. Similar to \eqn{HDJ1} relations
hold for the other two- and three-dimensional $\l$-deformed models we use in this paper.
It is likely that a similar relation, differing only on the numerical coefficient of the right hand side, is true for all $\l$-deformed coset CFTs $SO(n+1)/SO(n)$ and their analytic continuations.

\subsection{$\l$-deformed model based on the coset $SL(2,\mathbb{R})/U(1)$}

The various expressions for the fields for the $\l$-deformed $SL(2,\mathbb{R})/U(1)$ model can be derived from the one above by applying the analytic continuation
\begin{equation}
\label{AnalCont1}
 k \rightarrow - k \, , \qquad \a \rightarrow i \tau \, , \qquad \beta \rightarrow - i \rho \, .
\end{equation}
This operation transforms the metric to
\begin{equation}
\label{metric2}
 ds^2 = k \frac{1 - \l}{ 1 + \l} \big(   d\r^2 - \coth^2 \r \, d\tau^2 \big) + \frac{4 k \l}{1 - \l^2} \big(  \sinh\tau \, \coth\r \, d\tau + \cosh\tau \, d\r  \big)^2 \,
\end{equation}
and the dilaton to
\begin{equation}
\label{DilatonAdS2La}
  \Phi = - \ln \sinh\rho \, .
\end{equation}
The undeformed metric is obtained for $\l = 0$
\begin{equation}
 ds^2 = k \big(   d\r^2 - \coth^2 \r \, d\tau^2 \big) \ ,
\end{equation}
corresponding to the $SL(2,\mathbb{R})/U(1)$ exact CFT \cite{Witten:1991yr}.
We will denote the background corresponding to \eqn{metric2} and \eqn{DilatonAdS2La} as $CAdS_{2,\l}$.

\no
The frame that describes the deformed geometry now is
\begin{equation}
\label{FrameAdS2La}
 \mathfrak{e}^1 = \l_- \big(  \sinh\tau \, d\rho + \cosh\tau \, \coth\rho \, d\tau  \big) \, ,  \mathfrak{e}^2 = \l_+ \big(  \cosh\tau \, d\rho + \sinh\tau \, \coth\rho \, d\tau  \big)
\end{equation}
and it is associated to a metric with signature $(-,+)$. The symmetries \eqn{sym1} and \eqn{sym2} still hold with the
appropriate renaming (the symmetry acts via a complexification since $\tau\to \tau-i\pi/2$).
Also, as in footnote \ref{FrameSimp} the combinations $d(e^{-\Phi}  \mathfrak{e}^i)=0$,
for  $i=1,2$.

\no
The dilaton beta function simply acquires an overall minus sign compared to the one for the $CS^2_\l$ case, i.e.
\begin{equation}
\label{BetaAdS2La}
 \b^{\Phi}_{CAdS_{2,\l}} = R + 4 \nabla^2 \Phi - 4 \big(  \partial \Phi \big)^2 = - \n \ ,
\end{equation}
while the Ricci tensor and the dilaton satisfy the following relation
\begin{equation}
\label{HDJ2}
 R_{ab} + 2 \nabla_a \nabla_b \Phi = \m \, \d_{ab} \, .
\end{equation}
Note again that the Kronecker $\d_{ab}$ appears in the right hand side of this relation even though the signature of the space is  Minkowski. Hence, as \eqn{HDJ1} this is a frame dependent relation.

\subsection{$\l$-deformed model based on the coset $SL(2,\mathbb{R})/SO(1,1)$}

Like in the $CAdS_{2,\l}$ case the fields for the deformed $SL(2,\mathbb{R})/SO(1,1)$ can be obtained from those of the $CS^2_\l$ by the analytic continuation
\begin{equation}
\label{AnalCont2}
 k \rightarrow - k \, , \qquad \a \rightarrow \tau \, , \qquad \beta \rightarrow - i \rho \, .
\end{equation}
Under this transformation the line element becomes
\begin{equation}
 ds^2 = k \frac{1 - \l}{ 1 + \l} \big(   d\r^2 + \coth^2 \r \, d\tau^2 \big) + \frac{4 k \l}{1 - \l^2} \big(  \cos\tau \, d\r - \sin\tau \, \coth\r \, d\tau  \big)^2 \  ,
\end{equation}
whereas the dilaton is given by \eqn{DilatonAdS2La}.
The undeformed metric is found by setting $\l = 0$ which simply gives
\begin{equation}
 ds^2 = k \big(   d\r^2 + \coth^2 \r \, d\tau^2 \big) \, ,
\end{equation}
corresponding to the $SL(2,\mathbb{R})/SO(1,1)$ exact CFT \cite{Witten:1991yr}.
The symmetries \eqn{sym1} and \eqn{sym2} still hold with the
appropriate renaming.
We will denote the above background as $CH_{2,\l}$.

\no
The corresponding frame for the deformed line element now is
\begin{equation}
\label{FrameH2La}
 \mathfrak{e}^1 = \l_- \big(  \sin\tau \, d\rho + \cos\tau \, \coth\rho \, d\tau  \big) \, ,
 \quad\, \mathfrak{e}^2 = \l_+ \big(  \cos\tau \, d\rho - \sin\tau \, \coth\rho \, d\tau  \big)
\end{equation}
and it is associated to a metric with signature $(+,+)$. 
As in footnote \ref{FrameSimp} the combinations $d(e^{-\Phi}  \mathfrak{e}^i )=0$, $i=1,2$.
The dilaton beta function is the same as the one given for the $CAdS_{2,\l}$ case \eqn{BetaAdS2La}, i.e. $ \b^{\Phi}_{CH_{2,\l}}= \b^{\Phi}_{CAdS_{2,\l}}$.
Finally, the Ricci tensor and the dilaton now satisfy the relation
\begin{equation}
\label{HDJ3}
 R_{ab} + 2 \nabla_a \nabla_b \Phi = \m \, \eta_{ab} \ .
\end{equation}


\section{Solutions with geometry of the form $\cM_6 \times CS^2_\l \times CH_{2,\l}$}
\label{Sec3}
\def\vol{\textrm{Vol}}

We start our constructions by looking for solutions whose geometry is given by the direct product of a six dimensional manifold and the $\l$-deformed geometry $CS^2_\l \times CH_{2,\l}$. Thus the ansatz for the ten-dimensional line element is
\begin{equation}
 ds^2 = ds^2(\cM_6) + \big(  e^6 \big)^2 + \big(  e^7 \big)^2 + \big(  e^8 \big)^2 + \big(  e^9 \big)^2 \, ,
\end{equation}
where we take $e^6$ and $e^7$ to be identified with $\mathfrak{e}^1$ and $\mathfrak{e}^2$ of eq. \eqref{FrameS2La} respectively and $e^8$ and $e^9$ with $\mathfrak{e}^1$ and $\mathfrak{e}^2$ of eq. \eqref{FrameH2La}.  We will also complete the NS sector by considering a dilaton given by the sum of the corresponding $\l$-deformed spaces
\begin{equation}
 \Phi = - \ln (\sin\b \, \sinh\rho) \ .
\end{equation}
The above information allows one to get already an idea of the curvature of the space $\cM_6$. This can be done by re-writing the dilaton equation \eqref{DilatonEOM} as
\begin{equation}
 R_{\cM_6} + \b^{\Phi}_{CS^2_\l} + \b^{\Phi}_{CH_{2,\l}} = 0 \ ,
\end{equation}
which due to \eqn{HDJ1} and \eqn{HDJ3} implies for its Ricci scalar  that
\begin{equation}
\label{RicciM6}
 R_{\cM_6} = 0 \  .
\end{equation}

\subsection{Type-IIA solutions containing $AdS_2$ factors}
\label{TypeIIAsolAdS2}

By proposing an appropriate ansatz for the RR-sector we can obtain a series of type-IIA solutions with $AdS_2$ factors. In the following lines we describe each case separately. Let us consider backgrounds supported by the following set of RR fields
\begin{equation}
 \begin{aligned}
  & F_2 = 2 e^{-\Phi} \big(  c_1 e^7 \wedge e^9 + c_2 e^6 \wedge e^8 + c_3 e^7 \wedge e^8 + c_4 e^6 \wedge e^9 \big) \, ,
  \\[5pt]
  & F_4 = 2 e^{-\Phi} e^0 \wedge e^1 \wedge \big(  c_5 e^7 \wedge e^9 + c_6 e^6 \wedge e^8 + c_7 e^7 \wedge e^8 + c_8 e^6 \wedge e^9 \big) \, ,
 \end{aligned}
\end{equation}
with $c_1, \ldots , c_8$ being constants.
With this choice the expression for $F_2$ mixes the two $\l$-deformed backgrounds.  The expression for $F_4$ mixes the entire $\l$-deformed background with a two-dimensional part of $\cM_6$.
Using the observation in footnote \ref{FrameSimp},
we immediately see that the Bianchi identity \eqref{BianchisIIA} for $F_2$ is satisfied while that for $F_4$ implies that (provided that $F_4$ is not identically zero)
\begin{equation}
\label{vooll1}
 d \big(  e^0 \wedge e^1 \big) = 0 \ .
\end{equation}
Moreover, the first of the flux equations in \eqref{FluxesIIA} gives
\begin{equation}
\label{Cond1}
 c_1 c_5 + c_2 c_6 + c_3 c_7 + c_4 c_8 = 0 \ ,
\end{equation}
while from the last two we get
\begin{equation}
\label{vooll2}
 d \big(  e^2 \wedge e^3 \wedge e^4 \wedge e^5  \big) = 0 \ .
\end{equation}

The next step is to investigate the structure of the Ricci tensor on $\cM_6$ by  analyzing the Einstein equations. It turns out that the non-vanishing components of the \emph{symmetric} tensor
$\cT^{IIA}_{ab}$ in eq. \eqref{EinsteinIIA} are
\begin{equation}
\label{TtensorSol1}
 \begin{aligned}
  & \cT^{IIA}_{aa} = - \big(    c^2_1 + c^2_2 + c^2_3 + c^2_4 + c^2_5 + c^2_6 + c^2_7 + c^2_8 \big) \eta_{aa} \ , \;\qquad a = 0,1 \, ,
  \\[5pt]
  & \cT^{IIA}_{aa} = \big(    - c^2_1 - c^2_2 - c^2_3 - c^2_4 + c^2_5 + c^2_6 + c^2_7 + c^2_8 \big) \d_{aa} \ , \qquad a = 2,3,4,5 \, ,
  \\[5pt]
  & \cT^{IIA}_{aa} = \big(    c^2_1 - c^2_2 + c^2_3 - c^2_4 - c^2_5 + c^2_6 - c^2_7 + c^2_8 \big) \eta_{aa} \ , \;\;\;\;\qquad a = 6,7 \, ,
  \\[5pt]
  & \cT^{IIA}_{aa} = \big(    c^2_1 - c^2_2 - c^2_3 + c^2_4 - c^2_5 + c^2_6 + c^2_7 - c^2_8 \big) \eta_{aa} \ , \;\;\;\;\qquad a = 8,9 \ ,
  \\[5pt]
  & \cT^{IIA}_{67} = 2 \big(  c_2 c_3 + c_1 c_4 - c_6 c_7 - c_5 c_8  \big) \  ,
  \\[5pt]
  & \cT^{IIA}_{89} = 2 \big(  c_1 c_3 + c_2 c_4 - c_5 c_7 - c_6 c_8  \big) \  .
 \end{aligned}
\end{equation}
The first two lines  imply that the Ricci tensor on $\cM_6$ must have the following structure
\begin{equation}
\label{Cond2}
 \begin{aligned}
  & R_{ab} = - \big(    c^2_1 + c^2_2 + c^2_3 + c^2_4 + c^2_5 + c^2_6 + c^2_7 + c^2_8 \big) \eta_{ab} = - r_1 \eta_{ab} \ , \quad a, b = 0,1 \, ,
  \\[5pt]
  & R_{ab} = \big(    - c^2_1 - c^2_2 - c^2_3 - c^2_4 + c^2_5 + c^2_6 + c^2_7 + c^2_8 \big) \d_{ab} = r_2 \d_{ab} \  , \;\;\;\;\;\; a, b = 2,3,4,5 \ .
 \end{aligned}
\end{equation}
Clearly $r_1 > 0$, while the condition \eqref{RicciM6} implies
\begin{equation}
\label{Cond3}
 R_{\cM_6} = 2 c^2_5 + 2 c^2_6 + 2 c^2_7 + 2 c^2_8 - 6 c^2_1 - 6 c^2_2 - 6 c^2_3 - 6 c^2_4 = 4 r_2 - 2 r_1 = 0
 \ .
\end{equation}
Therefore, the curvature $r_2=r_1/2 >0$. Hence, we see that $\cM_6$ can be written as the direct product of two Einstein spaces, a two-dimensional one of negative curvature and a four-dimensional one of positive curvature. We will denote this
direct product structure as\footnote{
Superscripts, such as the ones below, in the symbol of an $n$-dimensional manifold, i.e. $\cM_n^\pm$,
indicate the sign of the corresponding curvature. If no superscript appears, then this curvature can assume either sign.}
\be
\label{m6m2m4}
\cM_6 = \cM^-_2 \times \cM^+_4\ .
\ee
The two conditions \eqn{vooll1}
and \eqn{vooll2} are obviously satisfied being the volume forms of the corresponding spaces.
Hence, we see the existence of an $AdS_2$ factor in the geometry. The positive part of the six-dimensional geometry $ \cM^+_4$ will not mix with the rest. It can be any regular
four-dimensional manifold with curvature $r_1/2$.

\no
Moreover, from the last four lines of the equation \eqref{TtensorSol1} we obtain the following conditions for the constants $c_1 , \ldots , c_8$
\begin{equation}
\label{Cond4}
\begin{split}
  & c^2_1 - c^2_2 + c^2_3 - c^2_4 - c^2_5 + c^2_6 - c^2_7 + c^2_8 = - \m \ ,
  \\
  & c^2_1 - c^2_2 - c^2_3 + c^2_4 - c^2_5 + c^2_6 + c^2_7 - c^2_8 = \m \ ,
  \\
  & c_2 c_3 + c_1 c_4 - c_6 c_7 - c_5 c_8 =0 \ ,
   \\
  &
   c_1 c_3 + c_2 c_4 - c_5 c_7 - c_6 c_8 = 0\ .
 \end{split}
\end{equation}

\no
In order to determine the solution we have to solve the constraints given in eqs \eqref{Cond1}, \eqref{Cond2} and \eqref{Cond4}. This is a system of seven quadratic equations with eight unknown parameters. 
Let us now restrict ourselves to the following cases:

\vskip .3 cm
\no
{\bf Solution 1:}  In this case the constants are

\begin{equation}
 \begin{aligned}
  & c^2_3 = \frac{r_1}{8} - \frac{\m}{2} \  , \qquad  c^2_4 = \frac{r_1}{8} + \frac{\m}{2} \  , \qquad c^2_5 = c^2_6 = \frac{3 r_1}{8} \  ,
  \\[5pt]
  & c^2_1 = c^2_2 = c^2_7 = c^2_8 = 0 \  .
 \end{aligned}
\end{equation}
Reality requires that $c^2_3 \geqslant 0$ which imposes the following bound on $r_1$
\begin{equation}
 r_1 \geqslant 4 \m \  .
\end{equation}
Hence, in the undeformed limit $\l=0$ ($\m\to 0$) is enough that the curvature is positive. However, in the
non-Abelian T-duality limit (described in detail in App. B) where $\l$ approaches unity and simultaneously $k\to \infty$ we have that $\m\to 2$. Then, the bound becomes $r_1 \geqslant 8$. Similar comments hold for the solutions 2 \& 3 below.

\vskip .3 cm
\no
{\bf Solution 2:} In this case we have that
\begin{equation}
 \begin{aligned}
  & c^2_3 =\frac{c^2_8}{3} = \frac{r_1 - \m}{8} \  , \qquad  c^2_4 = \frac{c^2_7}{3} = \frac{r_1 + \m}{8} \  ,
  \\[5pt]
  & c^2_1 = c^2_2 = c^2_5 = c^2_6 = 0 \  .
 \end{aligned}
\end{equation}
Again, in order for the solution to be real we must require that $c^2_3 \geqslant 0$ which imposes the bound
\begin{equation}
 r_1 \geqslant \m \  .
\end{equation}

\vskip .3 cm
\no
{\bf Solution 3:} Now the constants are
\begin{equation}
 \begin{aligned}
  & c^2_1 = c^2_2 = \frac{r_1}{8} \  , \qquad c^2_7 = \frac{3 r_1}{8} + \frac{\m}{2} \  , \qquad c^2_8 = \frac{3 r_1}{8} - \frac{\m}{2} \  ,
  \\[5pt]
  & c^2_3 = c^2_4 = c^2_5 = c^2_6 = 0 \  .
 \end{aligned}
\end{equation}
We need to impose that $c^2_8 \geqslant 0$ in order for the RR fields to be real. As a result
\begin{equation}
 r_1 \geqslant \frac{4 \m}{3} \  .
\end{equation}
Note that, even for $\l=0$ the RR-fields still remain non-vanishing. In that case, 
the four-dimensional transverse space corresponds to the exact CFT
$SU(2)/U(1)\times SL(2,\mathbb{R})/SO(1,1)$. Then, this space supports the embedding of
$ AdS_2 \times \cM^+_4$ to type-IIA supergravity. Similar comments hold for other solutions 
in this paper.

\vskip .3 cm
\no
{\bf Solution 4:} The constants are
\begin{equation}
 \begin{aligned}
  & c^2_1 = c^2_2 = \frac{\big(  r_1 - \m \big) \big(  4 \m - r_1 \big)}{8 r_1} \  , \qquad c^2_4 = \frac{2 r^2_1 - 5 r_1 \m + 4 \m^2}{4 r_1} \  ,
  \\[5pt]
  & c^2_5 = c^2_6 = \frac{\big(  r_1 - \m \big) \big(  2 r^2_1 - 5 r_1 \m + 4 \m^2 \big)}{8 r_1 \m} \, , \qquad c^2_7 = \frac{r_1 \big(  4 \m - r_1  \big)}{4 \m} \  ,
  \\[5pt]
  & c^2_8 = \frac{\big(  r_1 - \m \big)^2 \big(  4 \m - r_1 \big)}{4 r_1 \m} \, , \qquad c^2_3 = 0 \  .
 \end{aligned}
\end{equation}
Imposing that $c^2_1 , \ldots , c^2_8 \geqslant 0$  we obtain the condition
\begin{equation}
 \m \leqslant r_1 \leqslant 4 \m \  .
\end{equation}
Unlike the previous solutions, here the curvature is bounded from above as well.
In the conformal limit the curvature $r_1$ must vanish and thus the corresponding geometric six-dimensional space becomes flat. On the other hand, in the non-Abelian T-duality
 limit $2\leqslant r_1\leqslant 8$. Similar comments hold for the remaining solution below.

\vskip .3 cm
\no
{\bf Solution 5:} Now we have that
\begin{equation}
 \begin{aligned}
   & c^2_1 = c^2_2 = \frac{\big(  r_1 - \m \big) \big(  6 r^2_1 - 7 r_1 \m + 4 \m^2 \big)}{8 r_1 \m} \ , \qquad c^2_3 = \frac{\big(  r_1 - \m \big)^2 \big(  4 \m - 3 r_1 \big)}{4 r_1 \m} \  ,
  \\[5pt]
  & c^2_4 = \frac{r_1 \big(  4 \m - 3 r_1  \big)}{4 \m} \  , \qquad c^2_5 = c^2_6 = \frac{\big(  r_1 - \m \big) \big(  4 \m - 3 r_1 \big)}{8 r_1} \  ,
  \\[5pt]
  & c^2_7 = \frac{6 r^2_1 - 7 r_1 \m + 4 \m^2}{4 r_1} \, , \qquad c^2_8 = 0 \  .
 \end{aligned}
\end{equation}
Requiring $c^2_1 , \ldots , c^2_8 \geqslant 0$  we obtain that
\begin{equation}
\m \leqslant r_1 \leqslant \frac{4 \m}{3} \  .
\end{equation}
We note that for the solutions 1 \& 3 the signs of the non-vanishing constants $c_i$ are independent, while for the solutions 2, 4 \& 5 they are correlated via the equation \eqref{Cond1} and the last two in \eqn{Cond4}.

\subsection{Type-IIB solutions containing $AdS_3$ factors}
\label{TypeIIBsolAdS3}

We turn our attention to type-IIB solutions. We try for the self-dual RR-form the ansatz
\ba
  && f_5 = e^6 \wedge e^8 \wedge \big(  c_1 \, e^0 \wedge e^1 \wedge e^2 + c_2 \, e^3 \wedge e^4 \wedge e^5  \big) + e^6 \wedge e^9 \wedge \big(  c_3 \, e^0 \wedge e^1 \wedge e^2 + c_4 \, e^3 \wedge e^4 \wedge e^5  \big) \  ,
  \nonumber
\\
  && F_5 = 2 \, e^{- \Phi} \big(   1 + \star \big) f_5 \  ,
\ea
with $c_1 , \ldots , c_4$ being constants. The Bianchi equation \eqref{BianchisIIB} for $F_5$ implies that
\begin{equation}
 d \big(  e^0 \wedge e^1 \wedge e^2 \big) = d \big(  e^3 \wedge e^4 \wedge e^5  \big)
 = 0 \ .
\end{equation}
Taking this into account the flux equation \eqref{FluxesIIB} for $F_5$ is automatically satisfied. For the Einstein equations \eqref{EinsteinIIB} we find that the non-vanishing entries of the \emph{symmetric} tensor $\cT^{IIB}_{ab}$ are given by
\begin{equation}
\label{TtensorSol2}
 \begin{aligned}
  & \cT^{IIB}_{aa} = - \big(    c^2_1 + c^2_2 + c^2_3 + c^2_4 \big) \eta_{aa} \ , \;\qquad a = 0,1,2 \ ,
  \\[5pt]
  & \cT^{IIB}_{aa} = \big(    c^2_1 + c^2_2 + c^2_3 + c^2_4 \big) \d_{aa} \  , \;\;\;\;\qquad a = 3,4,5 \, ,
  \\[5pt]
  & \cT^{IIB}_{aa} = \big(    c^2_1 - c^2_2 + c^2_3 - c^2_4 \big) \eta_{aa} \  , \;\;\;\;\qquad a = 6,7 \, ,
  \\[5pt]
  & \cT^{IIB}_{aa} = \big(    c^2_1 - c^2_2 - c^2_3 + c^2_4 \big) \eta_{aa} \  , \;\;\;\;\qquad a = 8,9 \  ,
  \\[5pt]
  & \cT^{IIB}_{67} = 2 \big(  c_1 c_4 - c_2 c_3  \big) \  ,
  \\[5pt]
  & \cT^{IIB}_{89} = 2 \big( c_2 c_4 - c_1 c_3  \big) \  .
 \end{aligned}
\end{equation}
The first two lines imply that the Ricci tensor on $\cM_6$ must have the structure
\begin{equation}
\label{Cond5}
 \begin{aligned}
  & R_{ab} = - \big(    c^2_1 + c^2_2 + c^2_3 + c^2_4 \big) \eta_{ab} = - r \eta_{ab} \  , \quad a, b = 0,1,2 \  ,
  \\[5pt]
  & R_{ab} = \big(    c^2_1 + c^2_2 + c^2_3 + c^2_4 \big) \d_{ab} = r \d_{ab} \  , \;\;\;\;\;\; a, b = 3,4,5 \  .
 \end{aligned}
\end{equation}
Clearly $r > 0$, while the condition \eqref{RicciM6} is satisfied trivially. Hence we see that $\cM_6$ can be written as a direct product of a three-dimensional Einstein space of negative curvature and another one of positive curvature, i.e.
\be
\cM_6 = \cM^-_3 \times \cM^+_3 \ .
\ee
Obviously this geometry contains $AdS_3$ factors.

\no
The last four lines of eq. \eqref{TtensorSol2} imply the following constraints for the parameters $c_1 , \ldots , c_4$
\begin{equation}
\label{Cond6}
 \begin{aligned}
  & c^2_1 - c^2_2 + c^2_3 - c^2_4 = - \m \  ,
  \qquad c^2_1 - c^2_2 - c^2_3 + c^2_4 = \m \  ,
  \\[5pt]
  & c_1 c_4 - c_2 c_3 = 0 \ ,\qq
  c_2 c_4 - c_1 c_3 = 0 \ .
 \end{aligned}
\end{equation}
All in all we have to solve five constraints for the parameters $c_1, \ldots , c_4$. These are summarized in the equations \eqref{Cond5} and \eqref{Cond6}.
We can now distinguish the following two cases:

\vskip .3 cm
\no
{\bf Solution 1:} The constants are
\begin{equation}
 c_1 = c_2 = 0\ ,\qq
 c^2_3 = \frac{r - \m}{2} \, , \qquad c^2_4 = \frac{r + \m}{2} \  .
\end{equation}
Reality implies  that $c^2_3 \geqslant 0$ giving a lower bound for $r$, i.e.
\begin{equation}
 r \geqslant \m \ .
\end{equation}
The possible signs of $c_3$ and $c_4$  are all independent.

\vskip .3 cm
\no
{\bf Solution 2:} In this case $\l=0$ and the constraints are solved for
\begin{equation}
c_1 = \pm c_2 \ne 0\ ,\qq
 c^2_3 = c^2_4 = \frac{r}{2} - c^2_1 \  ,
\end{equation}
where now $c_1$ is arbitrary. Reality requires that $r\geqslant 2 c_1^2$.  Here the signs of the constants $c_i$ are correlated through the last two equations in \eqn{Cond6}.

\no
Note that in the above solutions the expression for the RR five-form completely mixes the
 three-dimensional subspaces of $\cM_6$ with the $\l$-deformed parts of the geometry.

\section{Solutions with geometry of the form $\cM_8 \times CS^2_\l$}
\label{Sec4}

Another interesting category of backgrounds is that with geometries written as direct products of an eight-dimensional manifold and a two-dimensional deformed space. This allows more room to obtain higher than three-dimensional $AdS$ spaces as part of the full supergravity solution.
In this section we will take the deformed part to be $CS^2_\l$, thus for the ten-dimensional geometry we make the ansatz  given by the line element
\begin{equation}
\label{mttr2}
 ds^2 = ds^2(\cM_8) + \big(  e^8 \big)^2 + \big(  e^9 \big)^2 \  ,
\end{equation}
where we take $e^8$ and $e^9$ to be identified with $\mathfrak{e}^1$ and $\mathfrak{e}^2$ of eq. \eqref{FrameS2La}, respectively. We also assume that the NS two-form vanishes while the dilaton is given by \eqref{DilatonS2La}.
From the dilaton equation \eqref{DilatonEOM} we get the following for the curvature of $\cM_8$
\begin{equation}
 R_{\cM_8} + \b^{\Phi}_{CS^2_\l} = 0 \  .
\end{equation}
Using \eqn{dilsul} this implies that $\cM_8$ is a constant negative curvature space
\begin{equation}
\label{Cond7}
 R_{\cM_8} = - \n \  .
\end{equation}
We will now study the geometric characteristics of $\cM_8$ by proposing specific ansatze for the RR fields.

\subsection{Type-IIB solutions containing $AdS_2$ \& $AdS_4$ factors}
\label{Sec41}

Let us focus on the case of type-IIB backgrounds whose RR sector consists of the following fields
\begin{equation}
\label{RRAdS24}
 F_1 = 2 e^{-\Phi} \big( c_1 \, e^8 + c_2 \, e^9 \big) \  , \qquad F_5 = 2 e^{-\Phi} \big(  1 + \star \big) e^4 \wedge e^5 \wedge e^6 \wedge e^7 \wedge \big( c_3 \, e^8 + c_4 \, e^9 \big) \  ,
\end{equation}
where the $c_1 , \ldots , c_4$ are taken to be constants. With this assumption all the Bianchi equations \eqref{BianchisIIB} are trivially satisfied except that for $F_5$ which implies that
\begin{equation}
\label{CondWedges1}
 d \big(  e^0 \wedge e^1 \wedge e^2 \wedge e^3 \big) = d \big(   e^4 \wedge e^5 \wedge e^6 \wedge e^7  \big) = 0 \  .
\end{equation}
This also ensures the validity of the flux equations \eqref{FluxesIIB}.

Next we compute the non-vanishing components of $\cT^{IIB}_{ab}$ which are
\begin{equation}
\label{TtensorSol3}
 \begin{aligned}
  & \cT^{IIB}_{aa} = - \big(  c^2_1 + c^2_2 + c^2_3 + c^2_4 \big) \eta_{aa} \  , \;\qquad a = 0,1,2,3 \, ,
  \\[5pt]
  & \cT^{IIB}_{aa} = \big(  - c^2_1 - c^2_2 + c^2_3 + c^2_4 \big) \d_{aa} \   , \qquad a = 4,5,6,7 \, ,
  \\[5pt]
  & \cT^{IIB}_{aa} = \big(  - c^2_1 + c^2_2 - c^2_3 + c^2_4 \big) \eta_{aa} \  , \qquad a = 8,9 \, ,
  \\[5pt]
  & \cT^{IIB}_{89} = 2 \big(  c_1 c_2 + c_3 c_4  \big) \  .
 \end{aligned}
\end{equation}
From the first two lines we can read the structure of the Ricci tensor on $\cM_8$
\begin{equation}
\label{Cond8}
 \begin{aligned}
  & R_{ab} = - \big(  c^2_1 + c^2_2 + c^2_3 + c^2_4 \big) \eta_{ab} = - r_1 \eta_{ab} \  , \;\;\;\;\qquad a, b = 0,1,2,3 \, ,
  \\[5pt]
  & R_{ab} = \big(  - c^2_1 - c^2_2 + c^2_3 + c^2_4 \big) \d_{ab} = r_2 \d_{ab} \  , \;\;\;\;\;\;\;\qquad a, b = 4,5,6,7 \  .
 \end{aligned}
\end{equation}
Notice that $r_1 > 0$. This tells us that $\cM_8$ can split into a direct product of two four-dimensional Einstein spaces one of which has negative curvature, i.e.
\be
\cM_8 = \cM^-_4 \times \cM_4\ .
\ee
Restricting ourselves to $AdS$ solutions we can think of two possibilities. One is to take $\cM^-_4$ to be the direct product of $AdS_2$ with $R_{ab} = - r_1 \eta_{ab}$ and a two-dimensional space of negative curvature $\cM^-_2$ with $R_{ab} = - r_1 \d_{ab}$ and Euclidean signature, e.g. $H_2$.
The second, perhaps more interesting, possibility is to take $\cM^-_4$ to be $AdS_4$ normalized such that $R_{ab} = - r_1 \eta_{ab}$. Moreover, the condition \eqref{Cond7} relates the curvatures $r_1$ and $r_2$ with the deformation parameter $\l$ as
\begin{equation}
\label{Cond7v2}
 r_1 - r_2 = \frac{\n}{4} \  .
\end{equation}
The last two lines of \eqref{TtensorSol3} can be thought of as constraints on the parameters $c_1, \ldots ,c_4$
\begin{equation}
\label{Cond9}
 c^2_1 - c^2_2 + c^2_3 - c^2_4 =  \m \  , \qquad c_1 c_2 + c_3 c_4 = 0 \ .
\end{equation}
Hence, in order to fully determine the solution one has to solve the conditions given in eqs. \eqref{Cond8}, \eqref{Cond7v2} and \eqref{Cond9} for the constants $c_1, \ldots ,c_4$ and the curvatures $r_1$ and $r_2$. The solution is
\begin{equation}
 \begin{aligned}
  & c_1 = s_1 \sqrt{\frac{\big(  \m + r_1 \big) \big(  \n + 4 \m - 4 r_1 \big)}{16 \m}} \ , \qquad  c_2 = s_2 \sqrt{\frac{\big(  \m - r_1 \big) \big(  \n - 4 \m - 4 r_1 \big)}{16 \m}} \ ,
  \\[5pt]
  & c_3 = s_3 \sqrt{\frac{\big(  \m + r_1 \big) \big(  4 r_1 + 4 \m - \n \big)}{16 \m}} \  , \qquad c_4 = s_4 \sqrt{\frac{\big(  \m - r_1 \big) \big(  4 r_1 - 4 \m - \n \big)}{16 \m}} \  ,
 \end{aligned}
\end{equation}
where $s_i = \pm 1$ satisfy the condition $s_1 \, s_2 + s_3 \, s_4 = 0$. This is a one-parameter family of solutions since the constants $c_1 , \ldots , c_4$ depend on $r_1$.

In order to ensure the reality of the solution we need to impose that all the arguments in the square roots of the previous expressions are positive. Thus we have the following possibilities:

\vskip .3 cm
\no
{\bf Range of  $\l \in \big(  0 , 4 - \sqrt{15}  \big]$:}
In that case the curvature $r_1$ is restricted inside the interval
\begin{equation}
 \frac{\n}{4} - \m \leqslant r_1 \leqslant \frac{\n}{4} + \m \ .
\end{equation}

\vskip .3 cm
\no
{\bf Range of $\l \in \big(  4 - \sqrt{15} , 1  \big]$:}
Now the curvature $r_1$ lies in the interval
\begin{equation}
 \m \leqslant r_1 \leqslant \frac{\n}{4} + \m \, .
\end{equation}
Here it is also possible to consider the non-Abelian T-duality limit.
In the above analysis we chose the arbitrary parameter to be $r_1$.

\no
When $\l \ne 4 - \sqrt{15}$ the parameters $c_3$ and $c_4$ do not vanish simultaneously and we have a mixing of the $\l$-deformed part with the four-dimensional subspaces of $\cM_8$, defined by the frames $(e^0, e^1, e^2, e^3)$ and $(e^4, e^5, e^6, e^7)$. This is done through a non-trivial RR five-form.

\no
In the special case where $\l = 0$ we obtain another solution
\begin{equation}
 \begin{aligned}
  & c_2 = s_2 \sqrt{\frac{1}{2k} - c^2_1} \, , \qquad c_3 = s_3 \sqrt{\frac{1}{2k} - c^2_1} \, , \qquad c_4 = s_4 \, c_1 \, ,
  \\[5pt]
  & r_1 = \frac{1}{k} \, , \qquad r_2 = 0 \, ,
 \end{aligned}
\end{equation}
with $s_{2,3,4} = \pm 1$ satisfying the relation $s_2 + s_3 \, s_4 = 0$. Here $c_1$ is a free parameter and the solution is real when $c^2_1 \leqslant 1/2k$.

\subsection{Type-IIB solutions containing $AdS_6$ factors}

Another solution of the type-IIB supergravity whose geometry supports an $AdS_6$ factor can be derived by the following ansatz
\begin{equation}
 F_1 = 2 e^{-\Phi} \big( c_1 \, e^8 + c_2 \, e^9 \big) \  , \qquad F_3 = 2 e^{- \Phi} e^6 \wedge e^7 \wedge \big(   c_3 \, e^8 + c_4 \, e^9  \big) \  ,
\end{equation}
where $c_1 , \ldots , c_4$ are constants. Obviously the Bianchi eq. \eqref{BianchisIIB} for $F_1$ is satisfied automatically, while that for $F_3$ gives the condition
\begin{equation}
 d \big(  e^6 \wedge e^7  \big) = 0 \  .
\end{equation}
After elaborating the flux equations \eqref{FluxesIIB} with the last equation in mind we derive a second constraint which reads
\begin{equation}
 d \big(  e^0 \wedge e^1 \wedge e^2 \wedge e^3 \wedge e^4 \wedge e^5  \big) = 0 \  .
\end{equation}
Moreover, from the first of eq. \eqref{FluxesIIB} we obtain a constraint for the parameters $c_1 , \ldots , c_4$ which reads
\begin{equation}
\label{Cond10}
 c_1 \, c_3 + c_2 \, c_4 = 0 \  .
\end{equation}

From the Einstein equations \eqref{EinsteinIIB} we compute the components of $\cT^{IIB}_{ab}$
\begin{equation}
\label{TtensorSol4}
 \begin{aligned}
  & \cT^{IIB}_{aa} = - \big(  c^2_1 + c^2_2 + c^2_3 + c^2_4 \big) \eta_{aa} \  , \;\;\;\;\qquad a = 0,1,2,3,4,5 \, ,
  \\[5pt]
  & \cT^{IIB}_{aa} = \big(  - c^2_1 - c^2_2 + c^2_3 + c^2_4 \big) \d_{aa} \  , \;\;\;\qquad a = 6,7 \  ,
  \\[5pt]
  & \cT^{IIB}_{aa} = \big(  - c^2_1 + c^2_2 - c^2_3 + c^2_4 \big) \eta_{aa} \  , \;\;\;\qquad a = 8,9 \  ,
  \\[5pt]
  & \cT^{IIB}_{89} = 2 \big(  c_1 c_2 + c_3 c_4  \big) \  .
 \end{aligned}
\end{equation}
The last suggests that the Ricci tensor on the manifold $\cM_8$ has the form
\begin{equation}
 \label{Cond11}
 \begin{aligned}
  & R_{ab} = - \big(  c^2_1 + c^2_2 + c^2_3 + c^2_4 \big) \eta_{ab} = - r_1 \eta_{ab} \  , \;\;\;\;\qquad a, b = 0,1,2,3,4,5 \  ,
  \\[5pt]
  & R_{ab} = \big(  - c^2_1 - c^2_2 + c^2_3 + c^2_4 \big) \d_{ab} = r_2 \d_{ab} \  , \;\;\;\;\;\;\;\qquad a, b = 6,7 \  .
 \end{aligned}
\end{equation}
As a result, the eight-dimensional space can be written as a direct product of a six-dimensional Einstein space of negative curvature and another two-dimensional Einstein space, that is
\be
\cM_8 = \cM^-_6 \times \cM_2\ .
\ee
Clearly one possibility that fits in this category is $ \cM_8 = AdS_6 \times S^2$.
Also, we can re-write the condition \eqref{Cond7} as
\begin{equation}
\label{Cond7v3}
 3 r_1 - r_2 = \frac{\n}{2} \ .
\end{equation}
Two more constraints on the parameters $c_1, \ldots ,c_4$ arise from the last two lines of eq. \eqref{TtensorSol4}. It turns out that these have exactly the same form as the ones in eq. \eqref{Cond9}. So in order to completely determine the background one has to solve eqs \eqref{Cond10}, \eqref{Cond11}, \eqref{Cond7v3} and \eqref{Cond9}.
It turns out that the solutions are the ones we present below:

\vskip .3 cm
\no
{\bf Solution 1:} The constants are
\begin{equation}
 \begin{aligned}
  & c_1 = s_1 \sqrt{\frac{\n + 4 \m}{12}} \ , \qquad c_4 = s_4 \sqrt{\frac{\n - 8 \m}{12}} \ , \qquad c_2 = c_3 = 0 \  ,
  \\[5pt]
  & r_1 = \frac{\n - 2 \m}{6} \ , \qquad r_2 = - \m \ ,
 \end{aligned}
\end{equation}
with $s_i = \pm 1$. Requiring that $c^2_4 \geqslant 0$ implies that $\l$ takes values in $[0, 4 - \sqrt{15}]$. Notice also that $r_2 \leqslant 0$, which means that $\cM_2$ has negative curvature.

\vskip .3 cm
\no
{\bf Solution 2:} There is also a second solution where now the various parameters are
\begin{equation}
 \begin{aligned}
  & c_2 = s_2 \sqrt{\frac{\n - 4 \m}{12}} \ , \qquad c_3 = s_3 \sqrt{\frac{\n + 8 \m}{12}} \ , \qquad c_1 = c_4 = 0 \  ,
  \\[5pt]
  & r_1 = \frac{\n + 2 \m}{6} \ , \qquad r_2 = \m \ ,
 \end{aligned}
\end{equation}
with $s_i = \pm 1$. In order to have real solutions we need to impose that $c^2_2 \geqslant 0$. Thus $\l$ must take values in the interval $[0, 2 - \sqrt{3}]$. In this case $r_2 \geqslant 0$ signaling to a positive curvature for $\cM_2$.

Notice that, the second solution has always a non-vanishing RR three-form.
For the first one this is also true except when $\l \ne 4 - \sqrt{15}$. The existence of a non-trivial RR three-form mixes the $\l$-deformed part with a two-dimensional subspace of $\cM_8$ spanned by $(e^6,e^7)$. The $AdS_6$ part, more generally $ \cM^-_6 $, stands on its own.


\section{The $\cM_8 \times CH_{2,\l}$ case}
\label{Sec5}

Another possible class of backgrounds could be those whose geometry is the direct product of an eight-dimensional manifold and $CH_{2,\l}$. In principle these can be constructed following the same method presented in the previous sections. The corresponding line element will have the form of \eqn{mttr2}, but now $e^8$ and $e^9$ are identified with $\mathfrak{e}^1$ and $\mathfrak{e}^2$ of eq. \eqref{FrameH2La}, respectively. Like in all the previous examples we will assume that the NS two-form vanishes while the dilaton is given by \eqref{DilatonAdS2La}.
From the dilaton equation \eqref{DilatonEOM} we get that
$\cM_8$ is a constant positive curvature space, i.e.
\begin{equation}
\label{Cond12}
 R_{\cM_8} + \b^{\Phi}_{CH_{2,\l}} = 0 \quad\Longrightarrow\quad  R_{\cM_8} = \n\ .
\end{equation}
The geometric properties of $\cM_8$ are determined depending on the specific ansatze for the RR fields to which we now turn.
We are going to consider a type-IIB background with non-trivial RR form fields given by
\begin{equation}
 F_1 = 2 e^{-\Phi} \big( c_1 \, e^8 + c_2 \, e^9 \big) \  , \qquad F_3 = 2 e^{- \Phi} e^0 \wedge e^1 \wedge \big(   c_3 \, e^8 + c_4 \, e^9  \big) \  ,
\end{equation}
where as before $c_1 , \ldots , c_4$ are constants. Obviously the Bianchi eq. \eqref{BianchisIIB} for $F_1$ is satisfied automatically, while that for $F_3$ gives the condition
\begin{equation}
 d \big(  e^0 \wedge e^1  \big) = 0 \  .
\end{equation}
More constraints can be obtained by looking at the flux equations \eqref{FluxesIIB}. The second and the third equation from \eqref{FluxesIIB} give
\begin{equation}
 d \big(  e^2 \wedge e^3 \wedge e^4 \wedge e^5 \wedge e^6 \wedge e^7  \big) = 0 \  ,
\end{equation}
where we already took into account the previous condition on $e^0 \wedge e^1$. The first of the flux equations leads to an algebraic constraint for the constants $c_1 , \ldots , c_4$ which turns out to be the same as in eq. \eqref{Cond10}.

Looking at the Einstein equations \eqref{EinsteinIIB} we compute the components of $\cT^{IIB}_{ab}$
\begin{equation}
\label{TtensorSol5}
 \begin{aligned}
  & \cT^{IIB}_{aa} = - \big(  c^2_1 + c^2_2 + c^2_3 + c^2_4 \big) \eta_{aa} \  , \;\;\;\;\qquad a = 0,1 \, ,
  \\[5pt]
  & \cT^{IIB}_{aa} = \big(  - c^2_1 - c^2_2 + c^2_3 + c^2_4 \big) \d_{aa} \  , \;\;\;\qquad a = 2,3,4,5,6,7 \  ,
  \\[5pt]
  & \cT^{IIB}_{aa} = \big(  - c^2_1 + c^2_2 + c^2_3 - c^2_4 \big) \eta_{aa} \  , \;\;\;\qquad a = 8,9 \  ,
  \\[5pt]
  & \cT^{IIB}_{89} = 2 \big(  c_1 c_2 - c_3 c_4  \big) \  .
 \end{aligned}
\end{equation}
From the last it is clear that the Ricci tensor on the manifold $\cM_8$ has the form
\begin{equation}
 \label{Cond13a}
 \begin{aligned}
  & R_{ab} = - \big(  c^2_1 + c^2_2 + c^2_3 + c^2_4 \big) \eta_{ab} = - r_1 \eta_{ab} \  , \;\;\;\;\qquad a, b = 0,1 \  ,
  \\[5pt]
  & R_{ab} = \big(  - c^2_1 - c^2_2 + c^2_3 + c^2_4 \big) \d_{ab} = r_2 \d_{ab} \  , \;\;\;\;\;\;\;\qquad a, b = 2,3,4,5,6,7 \  .
 \end{aligned}
\end{equation}
Obviously, the eight-dimensional space can be written as a direct product of a two-dimensional Einstein space of negative curvature and another six-dimensional Einstein space, that is
\be
\cM_8 = \cM^-_2 \times \cM_6\ .
\ee
It is clear that this structure allows the existence of an $AdS_2$ factor decomposing the eight-dimensional space as $\cM_8 = AdS_2 \times \cM_6$.
Also, we can re-write the condition \eqref{Cond12} as
\begin{equation}
\label{Cond12v2}
 3 r_2 - r_1 = \frac{\n}{2} \ .
\end{equation}
The parameters $c_1, \ldots ,c_4$ obey two more algebraic equations which come from the last two lines of eq. \eqref{TtensorSol5}. These are
\begin{equation}
 \label{Cond13}
 c^2_2 - c^2_1 + c^2_3 - c^2_4 =  \m \  , \qquad c_1 c_2 - c_3 c_4 = 0 \, .
\end{equation}
The background can be completely determined by solving the eqs \eqref{Cond10}, \eqref{Cond13a}, \eqref{Cond12v2} and \eqref{Cond13}.
It turns out that we can only have one real solution which is
\begin{equation}
 \begin{aligned}
  & c_2 = s_2 \sqrt{\frac{4 \m - \n}{12}} \ , \qquad c_3 = s_3 \sqrt{\frac{8 \m + \n}{12}} \ , \qquad c_1 = c_4 = 0 \  ,
  \\[5pt]
  & r_1 = \m \ , \qquad r_2 = \frac{2 \m + \n}{6} \ ,
 \end{aligned}
\end{equation}
for all independent signs of $s_{2,3}=\pm 1$.
The solution is real only when $c^2_2 \geqslant 0$ which restricts $\l$ in the interval $[2 - \sqrt{3} , 1]$. When $\l$ approaches its maximum value one can take the
non-Abelian T-duality  limit.

\section{Embedding the $\l$-deformed coset CFT $SO(4)/SO(3)$}
\label{Sec6}

Here we are going to apply our strategy in order to embed the $\l$-deformed coset CFT $SO(4)/SO(3)$ and its non-compact counterpart $SO(3,1)/SO(2,1)$ \cite{Demulder:2015lva} into the type-IIB supergravity.
We first summarize the field content of the $\l$-models that correspond to these spaces. The line element of the deformed $SO(4)/SO(3)$ CFT which we denote
as $CS^3_\l$, is given in terms of the following frame (see eq. (A.5) of  \cite{Demulder:2015lva})
\begin{equation}
 \begin{aligned}
  & \mathfrak{e}^1 = - \frac{2 \l_-}{\sqrt{\cA} \, \om^2_+ \, \om} \Big[  \om^2_+ \, \om^2 \, x \, dx - \om^2_+ \, y \, dy - \cA \, \om \, d\om  \Big] \, ,
  \\[5pt]
  & \mathfrak{e}^2 = - \frac{2 \l_+}{\sqrt{\cA \, \cD} \, \om^2_+ \, \om} \Big[  \cD \, \om^2_+ \, \om^2 \, dx + x \Big(  \om^2_+ \, y \, dy + \cA \, \om \, d\om  \Big)  \Big] \, ,
  \\[5pt]
  & \mathfrak{e}^3 = \frac{2 \l_+}{\sqrt{\cD} \, \om^2_+ \, \om} \Big(\om^2_+ \, dy - y \, \om \, d\om  \Big) \, ,
 \end{aligned}
\end{equation}
where we have defined for convenience the functions
\begin{equation}
 \cA = 1 - x^2 - y^2 \, , \qquad \cD = 1 - x^2 \, , \qquad \om_\pm = \sqrt{1 \pm \om ^2} \, .
\end{equation}
Notice that the above frame corresponds to a metric with signature $(+,+,+)$. The geometry is also supported by the non-trivial dilaton  (see eq. (A.6) of  \cite{Demulder:2015lva})
\begin{equation}
 \Phi = - \frac{1}{2} \ln \frac{8 \cA \om^2}{\om^4_+} \, ,
\end{equation}
whereas the NS two-form is zero.
Some key formulas allowing for a good ansatze for the RR sector of the type-II supergravity
are
\begin{equation}
 d \big(  e^{- \Phi} \mathfrak{e}^1  \big) = 0 \, , \qquad d \big(  e^{- \Phi} \mathfrak{e}^1 \wedge \mathfrak{e}^3  \big) = d \big(  e^{- \Phi} \mathfrak{e}^2 \wedge \mathfrak{e}^3  \big) = 0 \, .
\end{equation}
Moreover, the dilaton beta-function for this model reads
\begin{equation}
\label{Pggge1}
 \b^{\Phi}_{CS^3_\l} = R + 4 \nabla^2 \Phi - 4 \big(  \partial \Phi  \big)^2 = \frac{3}{4} \n \, .
\end{equation}
Another useful relation is
\begin{equation}
\label{Pggge2}
 R_{ab} + 2 \nabla_a \nabla_b \Phi = - \frac{\m}{2} \, \eta_{ab} \, ,
\end{equation}
where now $\eta_{ab}$ is the three-dimensional Minkowski metric with signature $(-,+,+)$.
Note that, as in the case of the two-dimensional $\l$-deformed coset spaces
this is a frame depended relation as well.

The non-compact version of $CS^3_\l$, which we will denote as $CH_{3,\l}$, can be obtained via an analytic continuation, namely
\begin{equation}
 k \rightarrow - k \, , \qquad \om \rightarrow i \om \, .
\end{equation}
As a result, the right hand sides of \eqn{Pggge1} and \eqn{Pggge2} flip signs.

Retuning to type-II supergravity we will look for backgrounds with geometry of the form $\cM_4 \times CS^3_\l \times CH_{3,\l}$ and line element given by
\begin{equation}
 ds^2 = ds^2(\cM_4) + \big(  e^4 \big)^2 + \big(  e^5 \big)^2 + \big(  e^6 \big)^2 + \big(  e^7 \big)^2 + \big(  e^8 \big)^2 + \big(  e^9 \big)^2 \, ,
\end{equation}
where we identify $e^4 \rightarrow \mathfrak{e}^1$, $e^5 \rightarrow \mathfrak{e}^2$, $e^6 \rightarrow \mathfrak{e}^3$, while $e^7$, $e^8$ and $e^9$ are identified with the analytic continuations
\footnote{
 After applying the analytic continuation we also assume a relabeling of the coordinates in order to distinguish the two spaces $CS^3_\l$ and $CH_{3,\l}$.
}
(which we do not reproduce here) of $\mathfrak{e}^1$, $\mathfrak{e}^2$ and $\mathfrak{e}^3$ respectively. We will also make the hypothesis that the NS sector consists of a dilaton, which is given by the sum of the dilatons of the $\l$-models $CS^3_\l$ and $CH_{3,\l}$.
Due to the fact that the dilaton beta functions for $CS^3_\l$  and $CH_{3,\l}$ come with opposite sign the dilaton equation \eqref{DilatonEOM} implies that
\begin{equation}
\label{Cond14}
 R_{\cM_4} = 0 \, .
\end{equation}
Focusing on type-IIB backgrounds we take as the only non-vanishing RR-field that
\begin{equation}
F_5 = 2 e^{-\Phi} \big(  1 + \star \big) e^0 \wedge e^1 \wedge \big( c_1 \, e^5 \wedge e^6 \wedge e^7 + c_2 \, e^4 \wedge e^8 \wedge e^9 \big) \  ,
\end{equation}
where $c_1, c_2$ are constants to be determined.
The Bianchi identity for $F_5$ \eqref{BianchisIIB} implies
\begin{equation}
 d \big(  e^0 \wedge e^1 \big) = d \big(   e^2 \wedge e^3  \big) = 0 \  .
\end{equation}
The above conditions ensure that the flux equation for $F_5$ \eqref{FluxesIIB} is also satisfied. This is expected since when the NS two-form or the RR three-form vanish, the flux equation for $F_5$ is equivalent to the Bianchi equation for $F_5$.

Moving on  to the Einstein equations \eqref{EinsteinIIB} we compute the components
\begin{equation}
\label{TtensorSol6}
 \begin{aligned}
  & \cT^{IIB}_{ab} = - \big(    c^2_1 + c^2_2 \big) \eta_{ab} \ , \;\qquad a, b = 0,1 \ ,
  \\[5pt]
  & \cT^{IIB}_{ab} = \big(    c^2_1 + c^2_2 \big) \d_{ab} \  , \;\;\;\;\qquad a, b = 2,3 \, ,
  \\[5pt]
  & \cT^{IIB}_{ab} = \big(    c^2_2 - c^2_1 \big) \eta_{ab} \  , \;\;\;\;\qquad a, b = 4,5,6 \, ,
  \\[5pt]
  & \cT^{IIB}_{ab} = \big(    c^2_1 - c^2_2 \big) \eta_{ab} \  , \;\;\;\;\qquad a, b = 7,8,9 \, .
 \end{aligned}
\end{equation}
The first two lines in eq. \eqref{TtensorSol6} suggest that the Ricci tensor of $\cM_4$ is
\begin{equation}
\label{Cond16}
 \begin{aligned}
  & R_{ab} = - \big(  c^2_1 + c^2_2 \big) \eta_{ab} = - r \eta_{ab} \  , \qquad a, b = 0,1 \  ,
  \\[5pt]
  & R_{ab} = \big(  c^2_1 + c^2_2 \big) \d_{ab} = r \d_{ab} \  , \;\;\;\;\;\;\;\qquad a, b = 2,3 \  .
 \end{aligned}
\end{equation}
As a result the four-dimensional space $\cM_4$ splits into a direct product of two two-dimensional Einstein spaces, one of which is of negative curvature and spanned by the directions $(e^0, e^1)$ while the second has positive curvature and is spanned by $(e^3, e^4)$. From now on we will denote this splitting as $\cM_4 = \cM^-_2 \times \cM^+_2$. Clearly one can choose $\cM^-_2$ to be $AdS_2$, i.e. $\cM_4 = AdS_2 \times \cM^+_2$. From eq. \eqref{Cond16} one can also check that eq. \eqref{Cond14} is indeed valid. Moreover, from the last two lines of eq. \eqref{TtensorSol6} we obtain one more constraint
\begin{equation}
\label{Cond17}
 c^2_1 - c^2_2 = \frac{\m}{2} \, .
\end{equation}
To fully determine the background we need to solve eqs \eqref{Cond16} and \eqref{Cond17}. Since the total number of independent equations that we need to solve is two and the number of parameters we have is three $(r, c_1, c_2)$ one of the parameters will be free. Choosing the free parameter to be $r$ the final solution is then
\begin{equation}
 c_1 = s_1 \sqrt{\frac{r}{2} + \frac{\m}{4}} \, , \qquad c_2 = s_2 \sqrt{\frac{r}{2} - \frac{\m}{4}} \, ,
\end{equation}
for all possible choices of the signs $s_{1,2} = \pm 1$. The solution is real provided $r \geqslant \m/2$.

\section{Concluding remarks}
\label{Sec7}

In this work we constructed solutions of the type-II supergravity based on the $\l$-deformed CFTs on $SU(2)/U(1)$, $SL(2,\mathbb{R})/U(1)$ and $SL(2,\mathbb{R})/SO(1,1)$. In all the examples studied, the geometry of the backgrounds allows the existence of $AdS_n$ factors with $n = 2,3,4$ and $6$ as part of the ten-dimensional geometry. In particular, the cases considered in this paper refer to geometries of the form $\cM_6 \times CS^2_\l \times CH_{2,\l}$, $\cM_8 \times CS^2_\l$ and $\cM_8 \times CH_{2,\l}$, where the manifolds $\cM_6$ and $\cM_8$ transverse to the deformed parts can split into direct products of Einstein spaces with constant curvatures. It turns out that the subspace which contains the time direction is of negative curvature and thus we can always choose it to be an $AdS$. More examples can be constructed in a similar manner including $CAdS_{2,\l}$ as part of the full geometry. Moreover, more solutions can be found through generalization of the ansatze considered in the main text. For example, in Sec. \ref{Sec41} one could look for backgrounds with RR sector of the form
\begin{equation}
 \begin{aligned}
  & F_1 = 2 e^{-\Phi} \big( c_1 \, e^8 + c_2 \, e^9 \big) \  , \qquad F_3 = 2 e^{- \Phi} e^6 \wedge e^7 \wedge \big(   c_3 \, e^8 + c_4 \, e^9  \big) \, ,
  \\[5pt]
  & F_5 = 2 e^{-\Phi} \big(  1 + \star \big) e^4 \wedge e^5 \wedge e^6 \wedge e^7 \wedge \big( c_5 \, e^8 + c_6 \, e^9 \big) \  ,
 \end{aligned}
\end{equation}
where $c_1 , \ldots c_6$ are constants. An exhaustive analysis of all possible solutions is beyond the scope of this paper.

Our solutions seem to be non-supersymmetric. We have checked the dilatino equation for the first solution in Sec. \ref{TypeIIAsolAdS2} as well as for the solution in Sec. \ref{TypeIIBsolAdS3}. Indeed, the variation of the dilatino vanishes only when the Killing spinor is trivial, suggesting that the supersymmetry is completely broken. 
We expect that this is the  case for the other solutions  we found in this paper.
The two-dimensional $\s$-models corresponding to our supergravity solutions can easily be made integrable. The reason is that
all factors, i.e. the $\l$-deformed, as well as the $AdS$ spaces, are. In addition, we may certainly choose  the rest of the factors, e.g. $\cM_4^+$ in \eqn{m6m2m4} to correspond to an integrable $\s$-model as well. However, the RR-fields may spoil integrability for the full string solution.

The reader might have noticed that we have not provided solutions with an $AdS_5$ factor, the reason being that we did not find an appropriate ansatz for the RR-fields. Perhaps this can be done by allowing warping to take place between the various
factors in the metric. Our type-IIA solutions can be trivially lifted to eleven-dimensional supergravity. Nevertheless,
it will be very interesting to find solutions of eleven-dimensional supergravity  with no type-IIA origin.

One might also wonder if the same method can be applied for deformed CFTs based on group spaces. The easiest option would be to try to embed the $\l$-model on $SU(2)$. This seems more challenging than the coset cases due to the
non-vanishing NS two-form. In a past attempt to embed these to supergravity led to solutions within the type-II*-theory
\cite{Sfetsos:2014cea} in which the RR-fields are purely imaginary. Perhaps progress
can be made by allowing warping between the different factors participating in the ansatz.

Another direction that would be interesting to follow, is to take advantage of these new solutions in order to bring the $\l$-deformed $\s$-models in contact with the ideas of the AdS/CFT correspondence. In the non-Abelian T-duality limit this opened new ways to understanding dual field theories,
e.g.  \cite{Sfetsos:2010uq,Itsios:2013wd,Lozano:2016kum,Lozano:2016wrs,Lozano:2017ole,
Itsios:2017cew,Itsios:2017nou}. Unlike the non-Abelian T-duality case here we have one-parameter families of supergravity solutions, thus making the use of the AdS/CFT correspondence more appealing.
In that respect, let's recall that one of the original motivations for constructing the $\l$-deformed $\s$-models was to
understand \cite{Sfetsos:2013wia} the global issues of supergravity backgrounds obtained by applying the non-Abelian T-duality
transformation. The way that the non-Abelian T-duality limit is taken, the originally
compact variables become non-compact, e.g. see Appendix B.
Our results are important steps forward in the quest to understand global issues in non-Abelian T-duality,
by using instead of the non-Abelian T-dual backgrounds the corresponding $\l$-deformed ones.

Finally, it will be very interesting to consider the plane-wave limit of our solutions,
taken in such a way that there is still a left over $\l$-dependence in the resulting solution.
We intent to address some of these ideas in the near future.

\subsection*{Acknowledgments}

We would like to thank H. Nastase and D.C. Thompson for useful remarks. 
The work of G.I. is supported by FAPESP grant 2016/08972-0 and 2014/18634-9.

\appendix

\section{The type-IIA and type-IIB supergravities}
\label{AppA}

Here we give a brief account of the equations of motion for the type-IIA and type-IIB supergravities.


\subsection{The type-IIA supergravity}

The type-IIA supergravity is described by the following ten-dimensional action in the string frame
\begin{equation}
\label{ActionIIA}
 S_{IIA} = \frac{1}{2 \kappa^2_{10}} \int\limits_{M_{10}} \sqrt{- g} \Bigg[  e^{-2 \Phi} \Bigg(  R + 4 \big(  \partial \Phi \big)^2 - \frac{H^2}{12}  \Bigg) - \frac{1}{2} \Bigg(  \frac{F^2_2}{2} + \frac{F^2_4}{4!} \Bigg) \Bigg] - \frac{1}{2} B \wedge dC_3 \wedge dC_3 \, ,
\end{equation}
where the RR fields can be written in terms of the NS potentials $B, C_1, C_3$ as
\begin{equation}
\label{BianchisIIA}
 H = dB \, , \qquad F_2 = dC_1 \, , \qquad F_4 = dC_3 - H \wedge C_1 \, .
\end{equation}
From the last we can easily verify the Bianchi identities
\begin{equation}
 dH = 0 \, , \qquad dF_2 = 0 \, , \qquad dF_4 = H \wedge F_2 \, .
\end{equation}
From the action \eqref{ActionIIA} one can derive the equation of motion for the dilaton, which is
\begin{equation}
\label{DilatonEOM}
 R + 4 \nabla^2 \Phi - 4 \big(  \partial \Phi \big)^2 - \frac{H^2}{12} = 0 \, .
\end{equation}
The equation of motion for the metric leads to the Einstein equations
\begin{equation}
\label{EinsteinIIA}
 \begin{aligned}
  & \cE^{IIA}_{\m\n} = R_{\m\n} + 2 \,\nabla_\m \nabla_\n \Phi - \frac{1}{4} \, \big(  H^2 \big)_{\m\n} - \cT^{IIA}_{\m\n} = 0 \, ,
  \\[5pt]
  & \cT^{IIA}_{\m\n} = e^{2 \Phi} \Bigg[  \frac{1}{2} \big(  F^2_2  \big)_{\m\n} + \frac{1}{12} \big(  F^2_4  \big)_{\m\n} - {1\ov 4}g_{\m\n} \Bigg(  \frac{F^2_2}{2} + \frac{F^2_4}{24} \Bigg)   \Bigg] \, .
 \end{aligned}
\end{equation}
Finally, the equations of motion for the potentials give us the flux equations below
\begin{equation}
\label{FluxesIIA}
 \begin{aligned}
  & d \Big(  e^{- 2 \Phi} \star H  \Big) - F_2 \wedge \star F_4 - \frac{1}{2} F_4 \wedge F_4 = 0 \, ,
  \\[5pt]
  & d \star F_2 + H \wedge \star F_4 = 0 \, ,
  \\[5pt]
  & d \star F_4 + H \wedge  F_4= 0 \, .
 \end{aligned}
\end{equation}


\subsection{The type-IIB supergravity}

The action for the type-IIB supergravity in string frame is
\begin{equation}
\label{ActionIIB}
 S_{IIB} = \frac{1}{2 \kappa^2_{10}} \int\limits_{M_{10}} \sqrt{- g} \Bigg[  e^{-2 \Phi} \Bigg(  R + 4 \big(  \partial \Phi \big)^2 - \frac{H^2}{12}  \Bigg) - \frac{1}{2} \Bigg(  F^2_1 + \frac{F^2_3}{3!} + \frac{F^2_5}{2 \cdot 5!} \Bigg) \Bigg] - \frac{1}{2} C_4 \wedge H \wedge dC_2 \, ,
\end{equation}
where the RR fields can be written in terms of the NS potentials $B, C_0, C_2, C_4$ as
\begin{equation}
 H = dB \, , \qquad F_1 = dC_0 \, , \qquad F_3 = dC_2 - C_0 H \, , \qquad F_5 = dC_4 - H \wedge C_2 \, .
\end{equation}
From the last we can easily verify the Bianchi identities
\begin{equation}
\label{BianchisIIB}
 dH = 0 \, , \qquad dF_1 = 0 \, , \qquad dF_3 = H \wedge F_1 \, , \qquad dF_5 = H \wedge F_3 \, .
\end{equation}
From the action \eqref{ActionIIB} one can derive the equation of motion for the dilaton which turns to be the same as in the type-IIA supergravity, i.e. it is given by the formula \eqref{DilatonEOM}. The equation of motion for the metric leads to the Einstein equations
\begin{equation}
\label{EinsteinIIB}
 \begin{aligned}
  & \cE^{IIB}_{\m\n} = R_{\m\n} + 2 \,\nabla_\m \nabla_\n \Phi - \frac{1}{4} \, \big(  H^2 \big)_{\m\n} - \cT^{IIB}_{\m\n} = 0 \, ,
  \\[5pt]
  & \cT^{IIB}_{\m\n} = e^{2 \Phi} \Bigg[  \frac{1}{2} \big(  F^2_1  \big)_{\m\n} + \frac{1}{4} \big(  F^2_3  \big)_{\m\n} + \frac{1}{96} \big(  F^2_5  \big)_{\m\n} - g_{\m\n} \Bigg(  \frac{F^2_1}{4} + \frac{F^2_3}{24} \Bigg)   \Bigg] \, .
 \end{aligned}
\end{equation}
Finally, the equations of motion for the potentials give us the flux equations below
\begin{equation}
\label{FluxesIIB}
 \begin{aligned}
  & d \Big(  e^{- 2 \Phi} \star H  \Big) - F_1 \wedge \star F_3 - F_3 \wedge F_5 =0\, ,
  \\[5pt]
  & d \star F_1 + H \wedge \star F_3 = 0 \, ,
  \\[5pt]
  & d \star F_3 + H \wedge \star F_5 = 0
  \\[5pt]
  & d \star F_5 - H \wedge  F_3 = 0 \, .
 \end{aligned}
\end{equation}

\section{The non-Abelian T-duality limit}
\label{AppB}

In the cases where the deformation parameter can approach the identity one can consider the non-Abelian T-duality limit. For the geometries $CS^2_\l$, $CAdS_{2,\l}$ and $CH_{2,\l}$ this limit is obtained by considering large values of $k$ after the following rescaling
\begin{equation}
\label{NATDexp}
 \tau = \frac{\hat{\tau}}{2 k} \, , \qquad \r = \frac{\hat{\r}}{2 k} \, , \qquad \a = \frac{\hat{\a}}{2 k} \, , \qquad \b = \frac{\hat{\b}}{2 k} \, , \qquad \l = 1 - \frac{1}{k} + \ldots
\end{equation}
To see this in practice we will consider the non-Abelian T-duality limit of the first solution in section \ref{TypeIIAsolAdS2}. The resulting background is a solution of the type-IIA supergravity with the following field content
\begin{equation}
\label{NATDsol1}
 \begin{aligned}
  & ds^2 = ds^2_{\cM_2} + ds^2_{\cM_4} + \frac{d\hat{\a}^2}{2 \hat{\b}^2} + \frac{1}{2} \Big(  d\hat{\b} + \frac{\hat{\a}}{\hat{\b}} d \hat{\a} \Big)^2 + \frac{d\hat{\tau}^2}{2 \hat{\r}^2} + \frac{1}{2} \Big(  d\hat{\r} - \frac{\hat{\tau}}{\hat{\r}} d \hat{\tau} \Big)^2 \, ,
  \\[5pt]
  & e^{- \Phi} = \frac{\hat{\b} \hat{\r}}{4} \, , \qquad F_2 = \frac{c_3}{4} \big(  \hat{\a} d \hat{\a} + \hat{\b} d \hat{\b} \big) \wedge d \hat{\tau} + \frac{c_4}{4 k^2} \big(  \hat{\tau} d \hat{\tau} - \hat{\r} d \hat{\r} \big) \wedge d \hat{\a} \, ,
  \\[5pt]
  & F_4 = \frac{1}{4} \, \sqrt{\frac{3 r_1}{8}} \, \textrm{Vol}(\cM_2) \wedge \Big[ d \hat{\a} \wedge d \hat{\tau} + \big(  \hat{\a} d \hat{\a} + \hat{\b} d \hat{\b}  \big) \wedge \big(  \hat{\r} d \hat{\r} - \hat{\tau} d \hat{\tau}  \big)  \Big]  \, ,
 \end{aligned}
\end{equation}
where $c_3 = \sqrt{\frac{r_1}{8} - 1}$ and $c_4 = \sqrt{\frac{r_1}{8} + 1}$. Notice that we rescaled the dilaton appropriately in order to absorb powers of $k$.

The above background can be obtained by performing an non-Abelian T-duality
 transformation along $S^2$ and $H_2$ of the following type-IIA solution
\begin{equation}
 \begin{aligned}
  & ds^2 = ds^2_{\cM_2} + ds^2_{\cM_4} + ds^2_{S^2} + ds^2_{H_2} \, ,
  \\[5pt]
  & F_2 = 2 c_3\, \textrm{Vol}\big(  H_2  \big) + 2 c_4\, \textrm{Vol}\big(  S^2  \big) + 2 \sqrt{\frac{3 r_1}{8}}\, \textrm{Vol}\big(  \cM_2  \big) \, , \quad F_4 = 2 \sqrt{\frac{3 r_1}{8}} \textrm{Vol}\big(  \cM_4  \big) \, ,
 \end{aligned}
\end{equation}
where $c_3$ and $c_4$ are given below eq. \eqref{NATDsol1}.

The non-Abelian T-duality
 limit can be applied in all solutions we found in the main text, provided the deformation parameter can approach identity for large values of $k$. However, we will not exhaust all the cases studied.




\end{document}